# Tracking Attosecond Electronic Coherences Using Phase-Manipulated Extreme Ultraviolet Pulses

Andreas Wituschek[1]*, Lukas Bruder[1,2]*, Enrico Allaria[3], Ulrich Bangert[1], Marcel Binz[1], Roberto Borghes[3], Carlo Callegari[3], Giulio Cerullo[4], Paolo Cinquegrana[3], Luca Gianessi[3], Miltcho Danailov[3], Alexander Demidovich[3], Michele Di Fraia[3], Marcel Drabbels[5], Raimund Feifel[6], Tim Laarmann[7,8], Rupert Michiels[1], Najmeh Sadat Mirian[3], Marcel Mudrich[9], Ivaylo Nikolov[3], Finn H. O'Shea[3], Giuseppe Penco[3], Paolo Piseri[10], Oksana Plekan[3], Kevin Charles Prince[3], Andreas Przystawik[7], Primož Rebernik Ribič[3], Giuseppe Sansone[1], Paolo Sigalotti[3], Simone Spampinati[3], Carlo Spezzani[3], Richard James Squibb[6], Stefano Stranges[11], Daniel Uhl[1] & Frank Stienkemeier[1,12]

[1]Institute of Physics, University of Freiburg, Hermann-Herder-Str. 3, 79104 Freiburg, Germany
[2]Chemical Physics, Lund University, Naturvetarvägen 16, 22362 Lund, Sweden.
[3]Elettra-Sincrotrone Trieste S.C.p.A., 34149 Basovizza, Trieste, Italy
[4]IFN-CNR and Dipartimento di Fisica, Politecnico di Milano, Piazza L. da Vinci 32, 20133 Milano, Italy
[5]Laboratory of Molecular Nanodynamics, Ecole Polytechnique Fédérale Lausanne, Route Cantonale, CH-1015 Lausanne, Switzerland
[6]Department of Physics, University of Gothenburg, Origovägen 6 B, SE-412 96 Gothenburg, Sweden
[7]Deutsches Elektronen-Synchrotron DESY, Notkestraße 85, 22607 Hamburg, Germany
[8]The Hamburg Centre for Ultrafast Imaging CUI, Luruper Chaussee 149, 22761 Hamburg, Germany.
[9]Department of Physics and Astronomy, Aarhus University, Ny Munkegade 120, DK-8000 Aarhus, Denmark
[10]Universita` degli Studi di Milano, Via Festa del Perdono 7, 20122 Milano, Italy
[11]University of Rome "La Sapienza", Piazzale Aldo Moro 5, 00185 Roma, Italy
[12]Freiburg Institute of Advanced Studies, University of Freiburg, Albertstraße 19, 79104 Freiburg, Germany

*Correspondence to: andreas.wituschek@physik.uni-freiburg.de, lukas.bruder@physik.uni-freiburg.de

November 11, 2020

## Abstract

The recent development of ultrafast extreme ultraviolet (XUV) coherent light sources bears great potential for a better understanding of the structure and dynamics of matter. Promising routes are advanced coherent control and nonlinear spectroscopy schemes in the XUV energy range, yielding unprecedented spatial and temporal resolution. However, their implementation has been hampered by the experimental challenge of generating XUV pulse sequences with precisely controlled timing and phase properties. In particular, direct control and manipulation of the phase of individual pulses within a XUV pulse sequence opens exciting possibilities for coherent control and multidimensional spectroscopy, but has not been accomplished. Here, we overcome these constraints in a highly time-stabilized and phase-modulated XUV-pump, XUV-probe experiment, which directly probes the evolution and dephasing of an inner subshell electronic coherence. This approach, avoiding any XUV optics for direct pulse manipulation, opens up extensive applications of advanced nonlinear optics and spectroscopy at XUV wavelengths.

**Introduction**

Coherent phenomena are an intriguing aspect of the quantum world. Their dynamics reveal rich information on a quantum system including ultrafast decay mechanisms and couplings to its environment[1]. To probe coherent dynamics in real time, specialized ultrafast methods are necessary which rely on interferometric measurements, mapping the oscillatory phase evolution of excited coherences[1]. This is most challenging for the electronic degrees of freedom where oscillation periods scale inversely with excitation energy and thus require extremely high timing stability in the atto- to zeptosecond range. Yet, to capture the full dynamical picture of a system, it is essential to track the evolution of all degrees of freedom including electronic coherences[2].

The extension of coherent time-resolved spectroscopy to XUV photon energies is highly desirable as it offers unprecedented site-specificity by accessing localized inner-shell states and opens the door to attosecond time resolution[3–5]. However, these experiments are extremely challenging due to the required ultra-high phase stability and the lack of phase-matching/cycling schemes necessary to isolate subtle coherence signals[6]. For these reasons, a real-time study of the evolution of an XUV electronic coherence has not been reported and only few examples of quantum beat and vibrational wave packet (WP) studies have been demonstrated, so far[7–9].

Coherent control in the XUV spectral range is a second field of high interest, which likewise relies on the preparation and probing of coherent states of matter. In a bichromatic approach, coherent control was achieved by controlling the relative delay between two XUV pulses[10]. Another, more general approach, relies on phase manipulation of the XUV pulses. Pulse shaping technology, only available in the infrared to ultraviolet spectral range, has enabled advanced control schemes, with applications in nonlinear optics and the steering of chemical reactions[11]. In the XUV, phase manipulation has been indirectly shown by varying the chirp[12] or polarization[13] of the driving field. However, direct and independent manipulation of the relative phase and delay of XUV pulses in a pulse sequence has not been accomplished so far.

In this work, we implement a phase modulation technique for XUV pulse sequences, which facilitates both flexible coherent control schemes, and advanced coherent nonlinear spectroscopy techniques. To this end, phase-locked XUV pulse pairs with full control over delay and relative phase were prepared at the FERMI free-electron laser[14] (FEL) using twin seeding[15] combined with phase-manipulated seed pulses (Fig. 1). The intense phase-locked ultraviolet (UV) seed pulse pairs are created with a highly stable interferometer based on a monolithic design[16], and are used to seed the FEL process by means of the high gain harmonic generation (HGHG) scheme. This leads to the emission of fully coherent XUV pulse pairs at a specific harmonic of the seed wavelength[17].

## Results

**Independent phase and timing control of XUV pulse pairs**

Previous experiments indicated the possibility of XUV phase manipulation through the seed pulse properties[12,15]. In the current work, we significantly advance this approach by implementing high precision, decoupled phase and timing control of XUV pulse pairs while avoiding the challenge of phase shaping at XUV wavelengths. For this purpose, two phase-locked acousto-optic modulators (AOMs) control the relative phase $\phi_{21} = \phi_2 - \phi_1$ of the seed pulses (Fig. 1). Upon HGHG, the imprinted phase transfers to a well-defined phase shift of $n\phi_{21}$ for the XUV pulses at the $n^{\text{th}}$ harmonic, enabling flexible phase manipulation (Fig. 2a), while the applied twin-seeding concept allows for even higher pump-probe timing control than advanced XUV split-and-delay units[18] and is not restricted to bichromatic pump-probe schemes[10].

The XUV phase control is demonstrated by manipulating the phase of XUV Ramsey-type interference fringes for photon energies up to 47.5 eV (Fig. 2b, c). The high quality of timing and phase control is directly reflected in the interferograms. The extremely stable seed laser interferogram indicates that the phase stability is only limited by fluctuations picked up during the HGHG process. Here, timing jitter between seed pulses and the electron bunch (≈ 42 fs RMS) leads to phase jitter of the produced XUV pulses due to the residual longitudinal energy chirp of the electron bunch[12,15] (see supplementary Fig. 1). However, when using echo-enabled harmonic generation, this noise source can be efficiently reduced, as in this process the spectrotemporal properties of the XUV pulses are significantly less sensitive to electron-bunch jitter than in HGHG[19]. Note, that our concept allows for the independent control of both timing and phase properties of the XUV pulses. Combined with multiple-pulse seeding, many pulse shaping and nonlinear spectroscopy applications can be realized now at XUV wavelengths.

**Tracking XUV electronic coherences in helium**

One type of applications of this concept are advanced coherent spectroscopy schemes based on phase-cycling concepts[20]. Coherent multipulse spectroscopy schemes (for example two-dimensional spectroscopy, transient grating spectroscopy) require selective, background-free detection of the weak nonlinear signals. Phase-cycling is an effective concept for highly selective detection which has, in the infrared (IR) to UV spectral range, drastically improved the detection sensitivity and paved the way for a plethora of nonlinear spectroscopy applications[21,22]. In the XUV, these experiments are extremely challenging, as they involve sequences of multiple phase-locked XUV pulses with independently controllable timing and phase parameters. Here, we demonstrate such an experiment by tracking the temporal evolution of attosecond electronic coherences with phase-modulated XUV pulse sequences.

The model system examined is the 1s² → 1s4p transition in helium (Fig. 3a). The first XUV pulse creates a coherent superposition of ground and excited state (electronic WP), denoted $|\psi(t)\rangle$. The second XUV pulse, delayed by $\tau$, projects this WP onto a stationary population state, which is probed by ionization of the 1s4p state with a near infrared (NIR) pulse, yielding the signal

$$S \propto \langle \psi(\tau)|1s4p\rangle = A(\tau)e^{i\phi(\tau)} \qquad (1)$$

where $A(\tau)$ denotes the amplitude and $\phi(\tau)$ the phase evolution of the WP. According to the 1s² → 1s4p transition energy $E$ = 23.74 eV, the signal would oscillate with a period of $h/E$ = 174 as, requiring an extremely high pump-probe timing stability of $\delta_\tau$ < 20 as (corresponding to an optical path stability of $\delta_{OP}$< 6 nm) for adequate sampling.

Here, we apply a specialized phase-cycling scheme to solve this problem. By combining phase modulation of both XUV pulses with phase-synchronous lock-in detection, we are able to downshift the signal oscillation period by a factor of > 50 (rotating frame detection[20]) and remove most of the phase jitter from the signal while improving the overall sensitivity through efficient lock-in amplification (details on the phase-modulation scheme and the rotating frame detection can be found in the Methods section). Fig. 3b shows the respective time-domain interferogram recorded for the helium excitation, exhibiting clean periodic oscillations of the induced attosecond electronic WP in excellent agreement with theory (Fig. 3c). The data quality is remarkable considering the applied low FEL energy (≤ 30 nJ) and the probed low atom density in the sample, which is several orders of magnitudes lower than in typical transient absorption experiments[23,24]. This shows the high signal-to-noise performance and sensitivity of the method, even with the challenging XUV wavelength conditions.

The signal quality furthermore allows for direct Fourier analysis to gain spectral information (Fig 3d), yielding a frequency spectrum with a signal to noise ratio of 10. Careful preparation of the electron bunch combined with the greatly reduced acquisition times due to rotating frame sampling allowed us to monitor the WP oscillations up to 700 fs, leading to a high spectral resolution in the Fourier domain (Gaussian RMS width $\sigma$ = 6 meV). This is even slightly better than achieved in XUV transient absorption experiments[23,24]. In addition, the Fourier transform approach combines spectral resolution with highly selective detection types, here demonstrated for mass-resolved photoion and energy-gated photoelectron detection. Furthermore, the phase-cycling lock-in detection scheme, enables highly efficient filtering of background signals in addition to mass/energy gating. As such, we were able to uncover WP signals that are up to a factor of 200 smaller than the background count rates inside the selected ion-mass/electron-energy windows.

At short delays where both seed pulses overlap temporally (<150 fs), twin-pulse seeding is expected to break down due to the nonlinear response of the density modulation initiating the FEL process to optical interference effects, causing shot-to-shot amplitude modulation of the FEL radiation. This explains the amplitude drop of our data at <150 fs. Interestingly, we can still observe a clear oscillation and a stable signal phase (not shown), indicating that information could be also gained from this region.

**Phase-resolved real-time dephasing of a Fano resonance**

The helium study served as a model for an unperturbed quantum system exhibiting long-lived electronic coherences and negligible dephasing. In argon, we investigate a different situation for the $3s^23p^6 \rightarrow 3s^13p^66p^1$ transition. The 6p valence orbital couples to the Ar$^+$ continuum via configuration interaction (Fig. 4a), opening-up an autoionizing pathway that introduces significant dephasing. Thereby, the phase shift between direct and autoionization pathways leads to characteristic Fano profiles[25]. Recently, a two-electron WP was fully reconstructed with the imprinted phase information in the Fano line shape[24]. Moreover, the scattering behavior of an electron WP in the vicinity of a Fano resonance has been investigated[26,27]. On the opposite, the dephasing of the WP in its quasi-bound state as it tunnels into the continuum has not been experimentally accessible, so far. Here, we provide a full characterization of this case by tracking directly the time evolution of the Ar 3s-6p electronic coherence and dissecting amplitude $A(\tau)$ and phase function $\phi(\tau)$ of the excited WP. Note, that $\phi(\tau)$ is here directly obtained from experimental data without the need of an iterative algorithm or a theoretical model (details in Supplementary Note 1).

Figure 4b shows the recorded time domain transients from which the complex-valued WP signal $S(\tau) = A(\tau)e^{i\phi(\tau)}$ is constructed. Here, the signal decay directly reflects the tunneling of the WP into the continuum. Its Fourier transform $S(\omega)$ is connected to the susceptibility of the sample[28] $\chi(\omega) \propto i\, S(\omega)$, hence we simultaneously obtain the absorption ($\propto Re[S(\omega)]$) and dispersion ($\propto -Im[S(\omega)]$) curves of the resonance (Fig. 4c). The absorptive part shows the typical Fano line shape (theoretical derivation in Supplementary Note 2 and Supplementary Fig. 2) in good agreement with synchrotron measurements[29]. Note that absorptive and dispersive parts of the susceptibility are independently retrieved in the measurement without requiring the Kramers-Kronig relation. Such full characterization of the susceptibility is particularly challenging in the XUV spectral range and can be beneficial in spectral regions where many resonances are congested, for instance at absorption edges[30].

From $\chi(\omega)$ we obtain the spectral phase $\phi(\omega)$. It resembles the characteristic of a damped driven harmonic oscillator (Fig 4c), implying the measurement's sensitivity to the bound character of the electron WP while it coherently oscillates in the Coulomb potential of the atomic core, in contrast to Refs. 26, 27. This is in accordance with the fact, that we track the electronic dephasing relative to the system's steady ground state. This dynamic has been so far not accessible due to the required high phase stability to probe the attosecond coherent oscillations between excited and ground state. In Refs. 26, 27 the electron WP generated at a Fano resonance was characterized relative to a reference continuum WP which makes these experiments sensitive to the continuum state of the generated WP. This explains the different shape of the spectral phase retrieved in these experiments. We note, that in the presented measurement, the uncertainty of the absolute phase is still ≈ 1 rad, but with appropriate calibration of the apparatus, uncertainties < 0.1 rad are feasible in future experiments (for more details on the phase calibration see Methods section).

## Discussion

To the best of our knowledge, resolving electronic coherences spanning over 28.51 eV (i.e. 145 as oscillation period) in a time-resolved experiment has not been achieved so far, which in the studied example enabled direct time-domain observation of the dephasing of an inner subshell - valence shell coherence. While electronic dephasing is studied in many examples in the visible, in the XUV range it may provide selective information about the coupling of a specific site to the environment or real-time information about intra- and inter-particle decay mechanisms. Furthermore, the phase-sensitive detection provides information about the spectral phase of electron WPs without the need to prepare a second, known reference WP in the system. This will be of advantage when studying the WP evolution in complex molecular and solid state systems.

In conclusion, we introduced a highly sensitive phase-cycling-based coherent spectroscopy technique at XUV wavelengths and tracked the attosecond time evolution of electronic WPs in a phase-resolved fashion. Scaling photon energies using echo-enabled harmonic generation[19] will allow localized core states to be addressed. In addition, the unprecedented sensitivity of our method provides ideal conditions for applications with tabletop high harmonic generation (HHG) sources. The generation of phase-locked XUV pulse pairs with HHG using phase-locked seeding has already been demonstrated[31]. Thus, an extension of our method to HHG sources is experimentally at hand. Our phase-cycling method combined with coherent XUV light sources, introduces a versatile XUV spectroscopy toolbox for the wide range application of specialized nonlinear spectroscopy schemes, so far only accessible at visible wavelengths. Specifically-designed detection protocols to address fundamental problems of ultrafast non-adiabatic dynamics[32], many-body phenomena[6], or the control of chemical reactions along complex energy landscapes may now be realized in the XUV regime.

## Methods

**FEL setup and sample preparation:**

The experiments were performed at the low-density matter (LDM) beamline endstation[33] at the Free-Electron Laser FERMI. We tuned the FEL photon energy to $h\nu = 23.74$ eV (5$^{th}$ harmonic) for helium, and 28.51 eV (6$^{th}$ harmonic) for argon. The FEL pulse duration estimated from the spectral bandwidth was 57 ± 4 fs and 45 ± 3 fs for the 5$^{th}$ and 6$^{th}$ harmonic respectively, which is in good agreement with the predictions from literature[34]. The helium beam was generated by supersonic expansion through a pulsed nozzle, yielding a density on the order of $10^{13}$ cm$^{-3}$ in the interaction region. The argon beam was effused from a gas cell, yielding a much lower density of $\approx 10^9$ cm$^{-3}$. In order to avoid saturation of the detection electronics and space charge effects, the FEL energies were adjusted using metal filters, yielding $E_{\text{pulse}} \leq 30$ nJ and $E_{\text{pulse}} \approx 10$ µJ per pulse for the helium and the argon study, respectively. The FEL repetition rate was 50 Hz. In order to provide similar parameters of the electron bunch for both seed pulses at all delays, the longitudinal beam properties should be as uniform as possible. This was achieved by reducing the compression of the electron bunch, yielding a sufficiently constant region along the bunch, allowing for a delay range of $\approx$ 1 ps. The XUV pulses were focused into the atomic beam to a diameter of 70 µm FWHM.

**Phase-modulated coherent time domain XUV spectroscopy:**

The seed laser pulses ($\lambda \approx 261$ nm, $\Delta t = 100$ fs FWHM) are split in a Mach-Zehnder interferometer (MZI), creating phase-locked pulse pairs with controllable delay $\tau$ (Fig. 1). Two AOMs are driven with continuous wave radio frequencies in phase-locked mode. Each driving frequency can be controlled with 0.12 Hz frequency and 380 µrad phase resolution. By manipulating frequency and phase of the driving fields high precision control of the relative phase $\phi_{21}$ between the seed pulses is thus achieved. We note that in comparison to established pulse-shaping technologies, this phase control approach can handle the required high seed pulse energies (> 10 µJ) and provides a high purity, artifact-free phase manipulation. This is crucial for the application in strongly nonlinear processes such as high harmonic generation, where any phase shaping distortion is amplified by the nonlinear process.

To generate the phase-controlled XUV pulse sequences, the phase-manipulated twin-seed pulses interact with a relativistic electron bunch in the modulator of the FEL. Thereby amplitude and phase of the twin-seed pulse are imprinted on the longitudinal phase space of the electron bunch. This transfers to an electron density modulation in a dispersive chicane, finally leading to the emission of an XUV pulse pair at a specific harmonic of the seed wavelength in subsequent undulators.

We use the phase modulation scheme developed by Marcus and coworkers[28], to perform phase-modulated coherent time domain XUV spectroscopy. Here, $\phi_{21}$ is incremented on a shot-to-shot basis, leading to a quasi-continuous modulation of $\phi_{21}$ at a frequency of $\Omega \approx 3$ Hz. Via the HGHG process, the relative phase of the XUV pulses transfers to $n\phi_{21}$, leading to a phase modulation at a frequency of $n\Omega$. The XUV pulse pair creates an electronic WP $|\psi\rangle$ between the ground state $|g\rangle$ and excited state $|e\rangle$. We monitor the temporal evolution of the WP by ionization out of the excited state. In the case of helium, this was realized with an additional synchronized NIR laser, while argon ions were obtained by autoionization. The evolution of the WP gives rise to delay-dependent oscillations in the ion yield $S(\tau) = A(\tau)\cos(\omega_{eg}\tau + \phi_0)$. Here $\omega_{eg}$ is the transition frequency from the ground to the excited state ($\omega_{eg} = 3.61 \times 10^{16}$ rad s$^{-1}$ for the excited helium resonance), and $\phi_0$ is a phase shift determined by the parameters of the WP. The phase modulation imparts an additional real-time modulation on the ion yield: $S(\tau, t) = A(\tau)\cos(\omega_{eg}\tau + \phi_0 - n\Omega t)$. Simultaneously, we collinearly trace the MZI with a narrowband 266 nm continuous-wave laser, which undergoes the same phase modulation and records all jitter inside the MZI, yielding the reference signal for lock-in detection: $R(\tau, t) = \cos(\omega_{ref}\tau - \Omega t)$, where $\omega_{ref} = 7.08 \times 10^{15}$ rad s$^{-1}$ is the optical frequency of the reference laser. Referencing the lock-in amplifier to a harmonic of $R$ (same harmonic as in the HGHG process) and demodulating $S$ therewith yields the in-phase (real component) $X(\tau)$ and in-quadrature components (imaginary component) $Y(\tau)$ of $S=X+iY$, from which the signal amplitude $A = \sqrt{X^2 + Y^2}$ and phase $\phi = \arctan(Y/X)$ are readily constructed. From this we reconstruct the complex-valued WP signal $S(\tau) = A(\tau) \exp(i\overline{\omega}_{eg}\tau + i\phi_0)$, with $\overline{\omega}_{eg} = \omega_{eg} - n\omega_{ref}$.

Note, that we detect here the attosecond oscillations $\omega_{eg}$ of the WP signal in a reference frame rotating at frequency $n\omega_{ref}$, a technique commonly referred to as rotating frame detection[20]. This leads to a downshift of the signal frequency: $\omega_{eg} \to \overline{\omega}_{eg}$ , which, for the case of the helium measurement ($\overline{\omega}_{eg} = 6.93 \times 10^{14}$ rad s$^{-1}$) corresponds to a downshift factor of 52. This explains the $2\pi/\overline{\omega}_{eg} \approx$ 9 fs period of the signal in Fig 3. b) and c) instead of a $2\pi/\omega_{eg}$ =174 as period and allows for much larger delay steps. Similar numbers apply for the argon measurment. Furthermore, all phase jitter arising from the MZI cancels efficiently during the demodulation process, which considerably improves the signal quality.

Spectral information is deduced by Fourier transforming the downshifted quantum oscillations $FT(S(\tau))$. Here the real and imaginary parts correspond to the absorptive and dispersive parts of the line shape, respectively[28]. The phase calibration of the apparatus was done with the helium measurements following the routine in Ref. 28. In between the helium and argon measurements, the delay zero position of the seed interferometer shifted due to a technical failure. The shift was inferred in the post processing from the seed laser Ramsey-type interference fringes which lead to a larger uncertainty than for the helium measurements, where time zero was deduced from seed laser autocorrelation measurements. This explains the relatively large uncertainty of the absolute phase mentioned in the main text. More details of the optical seed laser phase modulation setup and the harmonic lock-in demodulation scheme can be found in Refs. 16 and 28, respectively.

**Data availability**

The data that support the findings of this study are available from the corresponding authors upon reasonable request.

**Code availability**

Code used to process the data of this study is available from the corresponding authors upon reasonable request.

**Acknowledgements**

Funding by the Bundesministerium für Bildung und Forschung (BMBF, 05K16VFB), by the European Research Council (ERC) with the Advanced Grant "COCONIS" (694965) and by the Deutsche Forschungsgemeinschaft (DFG) IRTG CoCo (2079) is acknowledged. We gratefully acknowledge the support of the FERMI staff.

**Author contributions**

A.W., L.B. and F.S. conceived the experiment. A.W., L.B., P.C., A.D., M.D., I.N. and P.S. prepared the seed laser phase modulation setup. E.A., L.G., N.S.M., G.P., P.R.R., C.S., F.H.S. and S.Sp. prepared the FEL. A.W., R.B., C.C., M.D.F., R.M., O.P., R.J.S. and D.U. prepared the end-station. A.W., L.B., M.B., U.B., C.C., G.C., M.D., M.D.F., R.F., T.L., M.M., R.M., A.P., K.C.P., O.P., P.P., G.S., R.J.S., S.St., D.U., and F.S. performed the measurements. A.W. and L.B. wrote the manuscript with input from all other authors. A.W. and L.B. contributed equally to this work.

## Competing interests

The authors declare no competing interests.

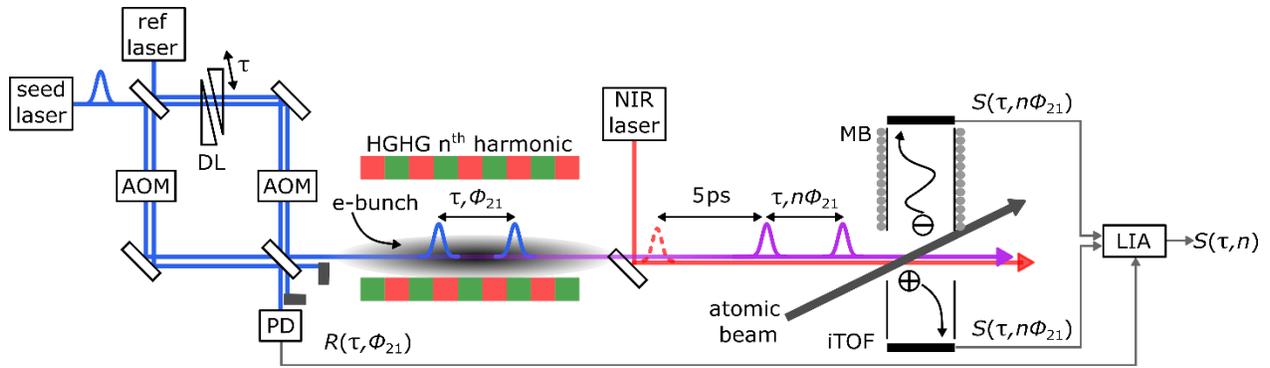

**Figure 1. Experimental scheme.** Intense twin-seed pulses are generated in an ultra-stable monolithic interferometer. The pulse delay $\tau$ is set by a wedge-based delay line (DL) and phase-locked acousto-optic modulators (AOMs) control the relative phase $\phi_{21}$ of the seed pulses. The time- and phase-controlled pulse pairs seed the high-gain harmonic generation (HGHG) process in the FEL, resulting in coherent XUV pulse pairs with precisely controlled timing and relative phase. The XUV pulse pair tracks the real time evolution of electronic coherences induced in an atomic beam sample. Detection is done via photoionization with a third pulse from a near infrared (NIR) laser. Both photoelectrons and -ions are detected with a combined magnetic bottle electron (MB) and ion time-of-flight (iTOF) spectrometer. A continuous-wave reference laser is used to trace the phase evolution and jitter in the interferometer, recorded with a photodiode (PD). This signal is used for rotating frame sampling and phase-sensitive detection of the mass/energy-gated ion/electron yields with a lock-in amplifier (LIA).

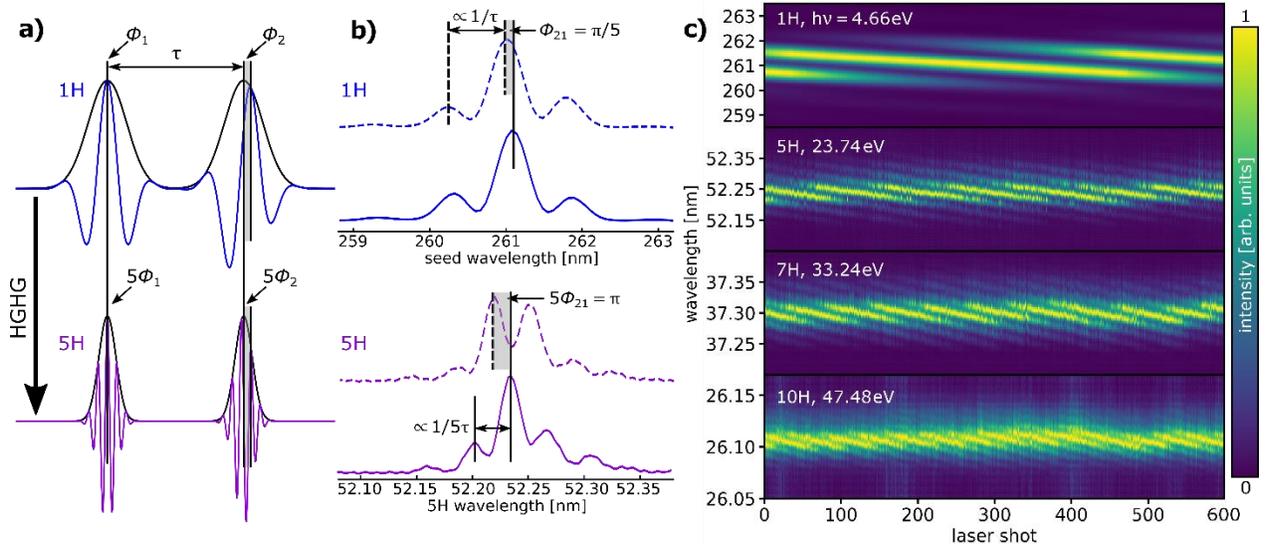

**Figure 2. XUV phase manipulation.** (**a**) Scheme of timing and phase control of the XUV pulse pair (violet) by manipulation of the seed pulse (blue) parameters. In this scheme, manipulation of pulse delay $\tau$ and phase difference $\phi_{21} = \phi_2 - \phi_1$ are decoupled, in contrast to previous work[10,12,15]. (**b**) Demonstration of XUV phase control in Ramsey-type interference fringes for fixed delay $\tau = 250$ fs and two different phase values $\phi_{21}$ recorded for the seed (blue) and its 5$^{th}$ harmonic (violet), respectively. The fringe spacing is inversely proportional to the pulse delay $\propto 1/\tau$, while the fringe phase is directly proportional to the phase difference $\propto \phi_{21}$. (**c**) Shot-to-shot phase manipulation of XUV pulses for different high harmonics for fixed delay $\tau = 250$ fs. Shown are Ramsey-type fringes of the seed (1H) and harmonic FEL pulses (5H-10H). Here, $\phi_{21}$ was incremented by 15 mrad steps in-between each laser shot leading to a quasi-continuous phase sweep. Full phase rotations of several periods of $2\pi$ are demonstrated without modification of the relative pulse delays. At the 10$^{th}$ harmonic, the spectrometer resolution is on the order of the fringe spacing, compromising the data quality slightly. All Ramsey-type fringes correspond to normalized single-shot events with no additional data processing applied.

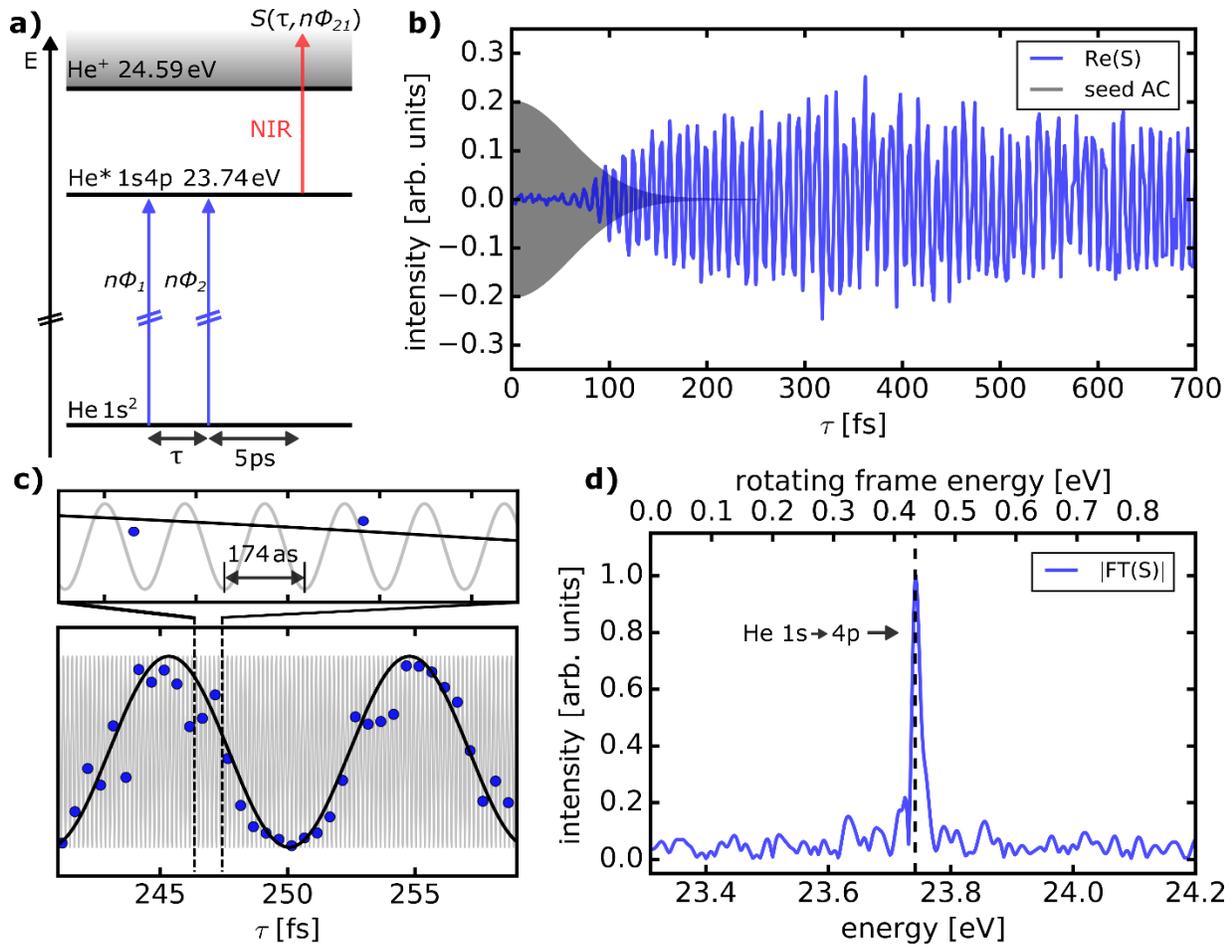

**Figure 3. XUV electronic coherence in helium.** (**a**) Energy scheme of helium with relevant levels and interaction pathways. (**b**) Real part of the downshifted WP signal, with the delay incremented in 2 fs steps. Temporal overlap region of seed pulses is indicated (grey area). Each data point corresponds to only 640 consecutive FEL shots, no additional data filtering was applied. (**c**) Data points taken with delay increment of 500 as (blue) compared to downshifted oscillation frequency of the excited transition (black). For comparison, the grey curve shows the theoretical rapid oscillation, that would be obtained without rotating frame detection ($T = 174$ as, see also inset). Due to rotating frame detection, we get a frequency downshift by a factor of $\approx 52$, transferring the attosecond beats to the femtosecond regime. (**d**) Fourier transform (absolute value) of the signal, showing the He 1s → 4p resonance. The upper energy scale shows the downshifted frequency range

obtained by rotating frame detection, the lower energy scale shows corresponding absolute energy values.

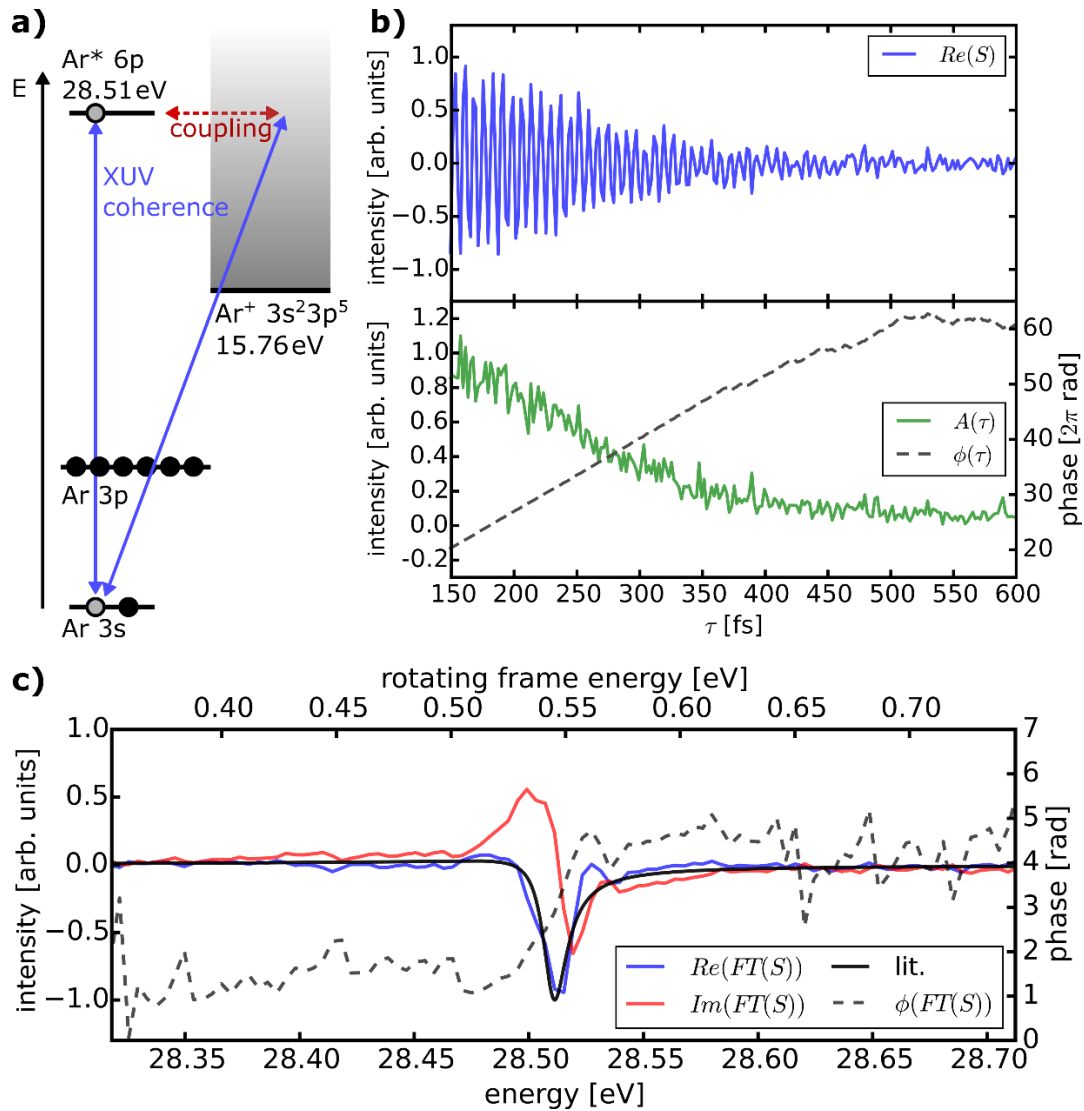

**Figure 4. Phase-resolved real-time dephasing of a Fano resonance in argon.** (**a**) Excitation scheme for preparation and probing of the Ar 3s-6p inner subshell-valence coherence. (**b**) Decay of the 3s-6p coherence in the time domain, with real part of the complex signal $S = A(\tau)e^{i\phi(\tau)}$ (blue) and its decomposition into amplitude $A(\tau)$ (green) and phase $\phi(\tau)$ (red). (**c**) Linear susceptibility of the process obtained from Fourier transform of the signal: Absorption (real part, blue) and dispersion (imaginary part, red) curve compared to the Fano lineshape from Ref. 29 (black) and spectral phase $\phi(\omega)$ (dashed grey), exhibiting a characteristic phase jump at the resonance of $\approx \pi$.

# Supplementary information

# Tracking Attosecond Electronic Coherences Using Phase-Manipulated Extreme Ultraviolet Pulses

A. Wituschek et. al

**Supplementary Note 1: Phase retrieval using heterodyned detection:**

The full characterization of an electron WP in amplitude and phase is a non-trivial problem and a major goal in attosecond metrology. The issue of dissecting the amplitude $A(t)$ and phase function $\phi(t)$ of an arbitrary signal $S(t) = A(t)e^{i\phi(t)}$ is well-known from e.g. optical pulse characterization, where many methods have been developed to solve this problem. However, it is difficult to apply these methods in the XUV spectral range, which makes the phase retrieval of WPs excited/probed in the XUV domain generally a challenging task.

We solve this problem by introducing heterodyned detection with a known reference waveform, a common method in signal processing. Instead of using optical heterodyning, which is challenging at XUV wavelengths, the imprinted phase modulation effectively shifts the signal down to the low kHz-frequency regime where standard lock-in electronics can be used for heterodyned detection with a known electronic waveform. The quadrature demodulation with the known reference then yields in-phase and in quadrature signal components from which amplitude and phase are readily reconstructed (see also Methods section).

We note that our approach for retrieving amplitude and phase of a WP is universal and does not require energy-resolved detection. This permits its implementation with arbitrary detection types like ion time-of-flight detection, velocity map imaging or reaction microscopes.

**Supplementary Note 2: Contribution of the Fano profile to the ion/electron count rate:**

In quantum interference experiments, typically the pathway interference between the pump excitation and the delayed probe excitation is probed. For the presented study of a Fano resonance, the situation is slightly more complex, as each pulse can excite two coherent pathways leading to the same final state (labeled $i = 1, 2$ in supplementary Fig. 2a). For pathway amplitudes $A_i$ the ion/electron signal is accordingly

$$S \propto |A_1(t) + A_2(t) + A_1(t+\tau) + A_2(t+\tau)|^2 . \tag{1}$$

Here, the interference term of pathway $i$ excited by the pump with pathway $j$ excited by the delayed probe ($i, j = 1, 2, i \neq j$) contributes the phase difference between $A_1$ and $A_2$ to the signal which is hence encoded in the measured ion/electron count rates.

For a quantitative derivation of the signal we apply time-dependent perturbation theory and adapt the calculations from supplementary Ref. 1 to a one-dimensional quantum interference experiment. The experimental observable is the ion/electron yield which is proportional to the population probability of the quasi-bound eigenstates $|k\rangle$ of the diagonalized Hamiltonian (supplementary Fig. 2b). Expanding the signal to second-order in the optical field and assuming delta-like excitation pulses yields:

$$S(\tau) \propto \int dk \, i |\mu_{kg}|^2 \Theta(\tau) e^{i(\omega_{kg} - i\gamma_{kg})\tau} , \tag{2}$$

where $\mu_{kg} = \mu_{cg}(\epsilon + q)/(\epsilon + i)$ denotes the $|g\rangle \to |k\rangle$ transition dipole moment with Fano parameter $q$ and reduced energy $\epsilon = (\omega_{kg} - \omega_{eg})/\gamma_e$. It is $\omega_{jg}$ the $|g\rangle \to |j\rangle$ transition frequencies, $\gamma_{kg}$ the dephasing rate of the $k$-$g$ coherence, $\gamma_e \propto V^2$ the dissipation/tunneling rate from the $|e\rangle$ state into the continuum and $\Theta(t)$ the Heaviside step function. For simplicity we omitted the phase modulation term.

A Fourier transform yields the spectral signal

$$S(\omega) \propto \int dk \, i |\mu_{kg}|^2 \frac{1}{\omega - \omega_{kg} - i\gamma_{kg}} . \tag{3}$$

For the dilute gas-phase sample probed in our study, we can assume $\gamma_{kg} \ll \gamma_e$. The real part of supplementary Eq. 3 is then

$$Re\{S(\epsilon, q)\} \propto \frac{(q + \epsilon)^2}{\epsilon^2 + 1} , \tag{4}$$

which describes the well-known Fano profile.

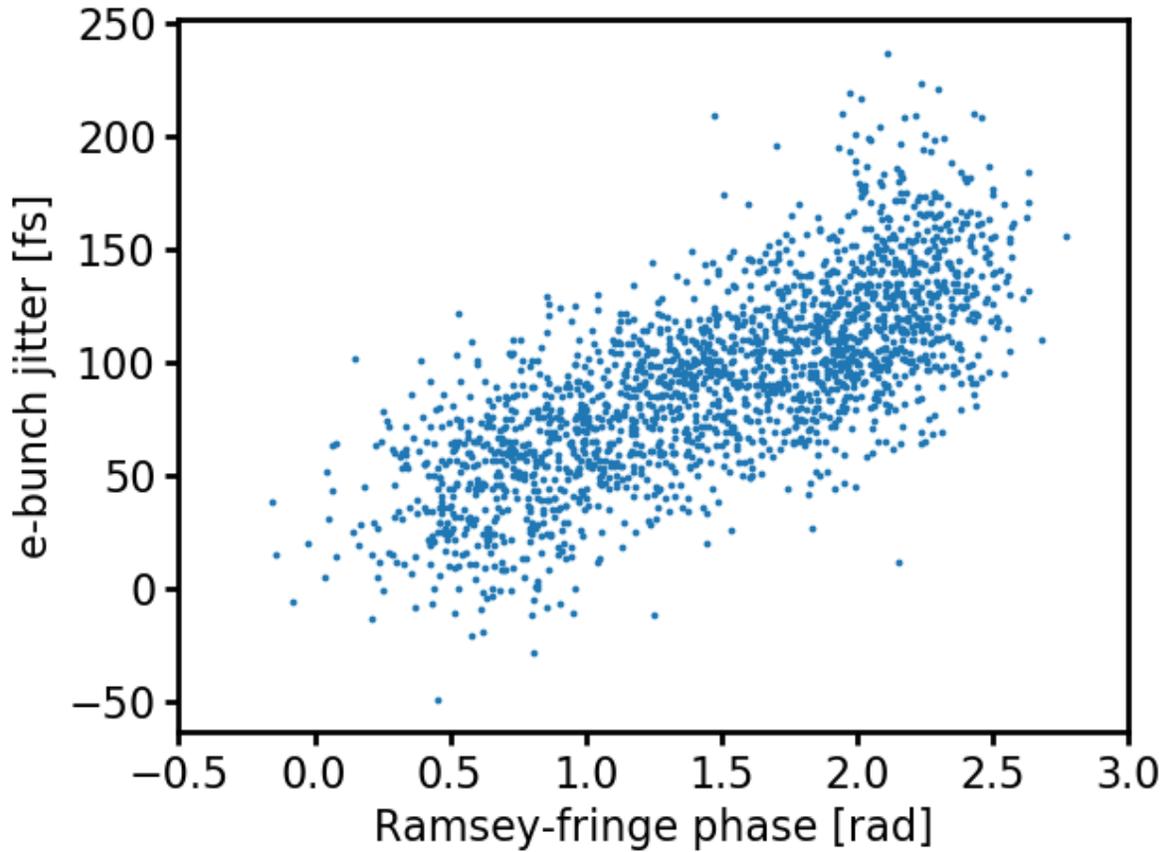

**Supplementary Figure 1** Phase stability of the fifth harmonic for 2000 consecutive shots. A strong linear correlation, similar to the one in suppementary Ref. 2, is observed between the phase of the Ramsey-type fringe pattern on the beamline spectrometer and the timing jitter between seed pulses and electron-bunch. The electron bunch jitter has an RMS value of 42 fs and the phase jitter of 0.6 rad in this measurement.

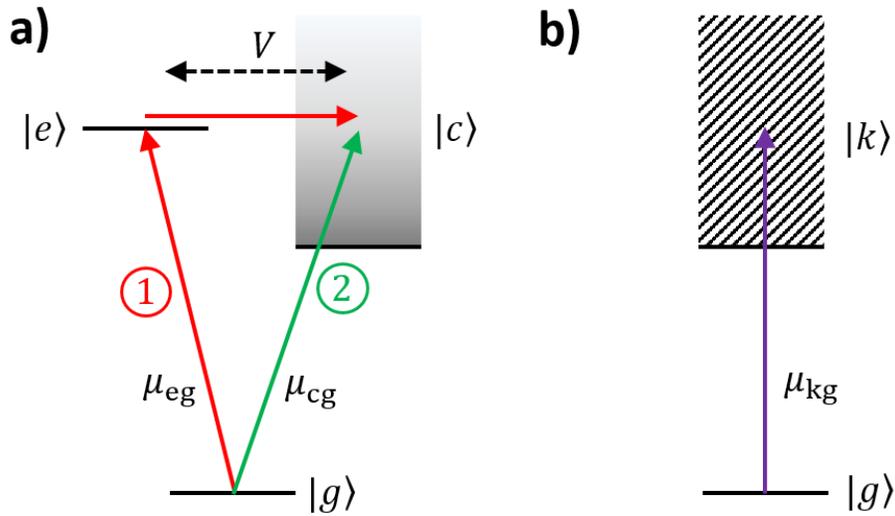

**Supplementary Figure 2** Level diagram of a model Fano system in a) with $|g\rangle$ ground, $|e\rangle$ bound excited and $|c\rangle$ continuum states and $\mu_{eg}, \mu_{cg}$ corresponding transition dipole moments. $V$ denotes the coupling between $|e\rangle$ and $|c\rangle$ states. b) shows the equivalent diagonalized system with $|g\rangle$ ground and $|k\rangle$ quasi-bound eigenstates of the Hamiltonian and $\mu_{kg}$ respective transition dipole moment.

**Supplementary References:**